# Very fast formation of superconducting $MgB_2$/Fe wires with high $J_c$

X.L. Wang, S. Soltanian, J. Horvat, M.J. Qin, H.K. Liu and S.X. Dou

Institute for Superconducting and Electronic Materials, University of Wollongong, Northfields Ave, NSW 2522 Australia

## Abstract

In this paper we have investigated the effects of sintering time and temperature on the formation and critical current densities of Fe-clad $MgB_2$ wires. $MgB_2$ wires were fabricated using the powder-in-tube process and sintered for different periods of time at predetermined temperatures. All the samples were examined using XRD, SEM and magnetisation measurements. In contrast to the common practice of sintering for several hours, the present results show that there is no need for prolonged heat treatment in the fabrication of Fe-clad $MgB_2$ wires. A total time in the furnace of several minutes is more than enough to form nearly pure $MgB_2$ with high performance characteristics. The results from $T_c$, $J_c$ and $H_{irr}$ show convincingly that the samples which were sintered for 3 minutes above 800 $^oC$ are as good as those sintered for longer times. In fact, the $J_c$ field performance for the most rapidly sintered sample is slightly better than for all other samples. $J_c$ of $4.5\times10^5$ A/cm$^2$ in zero field and above $10^5$ A/cm$^2$ in 2T at 15 K has been achieved for the best Fe-clad $MgB_2$ wires. As a result of such a short sintering there is no need for using high purity argon protection and it is possible to carry out the heat treatment in a much less protective atmosphere or in air. These findings substantially simplify the fabrication process, making it possible to have a continuous process for fabrication and reducing the costs for large-scale production of $MgB_2$ wires.

Key words: Magnesium boride, fast fabrication, superconducting wire, critical current

## 1. Introduction

Impressive progress has been made in the fabrication of $MgB_2$ wires since the discovery of superconductivity at 39 K in this material [1]. A number of techniques have been developed to improve the processing parameters for achieving high critical current densities [2-12]. Among these, the powder-in-tube (PIT) process appears most promising and practical. Some metals and alloys have been found to be suitable for sheath materials in the PIT process. Iron and its alloys in particular have been found to be not only non-poisoning to $MgB_2$ [6-10] but also capable of providing magnetic screening to reduce the effect of external applied magnetic fields on the critical currents [7,9]. Very high transport critical current densities have been reported for Fe and Fe alloy clad-$MgB_2$ wires by some groups [7-9, 11, 12].

The precursor powder is another important factor in the PIT process for the fabrication of high temperature superconductor wires. There are two different possible starting powders in the PIT process for fabrication of $MgB_2$ wires: fully reacted $MgB_2$ powder and a mixture of unreacted Mg + 2B powder. A high transport $J_c$ on the order of $10^4$-$10^5$ A/cm$^2$ at 20 K and 4.2 K has been reported for Cu/Fe/MgB2 tapes where reacted MgB2 powders were used as the core conductor and sintered at 900-1000 $^o$C for 0.5 h out of a total heat treatment time of more than 3 h including the initial heating [8]. By using unreacted Mg+2B powders and sintering at 800 $^o$ C for 1 h, we have successfully fabricated Fe clad $MgB_2$ tapes with a very high transport $J_c$ above $10^4$ A/cm$^2$ at 30 K and 1 Tesla and $I_c$ greater than 150 A [9].

Heat treatment has been applied in most PIT processes used for the fabrication of $MgB_2$. Based on all the reported data, the periods used for fabricating $MgB_2$ pellets and wires/tapes range from an hour to more than 48 h at sintering temperatures above 600-1000 $^o$C. All the heat treatments must be under high purity Ar protection from oxidation. High $J_c$ metal-sheathed $MgB_2$ tapes without any heat treatment were recently reported [4]. This process has advantages over those with heat treatment as it substantially simplifies the process and hence reduces the cost for wire fabrication. However, the sheath materials need to be of very high hardness in order to densify the $MgB_2$ core. High toughness metal is easily broken during the cold drawing and rolling process, and very careful and delicate design and control are mandatory [11,13]. Furthermore, as $MgB_2$ is very brittle, it would be a formidable task to overcome the crack problem for long length production without heat treatment.

In order to further improve and simplify the fabrication processes we have carried out a systematic study on the effect of sintering time on $MgB_2$ formation and $J_c$. In this paper, we present our findings that Fe-clad $MgB_2$ tapes with high $J_c$ can be sintered in a very short period, with a total time of several minutes in a furnace. This is highly relevant to the industrial fabrication and applications of MgB2 because it allows for continuous processing in a less or even non-protective environment.

## 2. Experimental Details

Powders of magnesium (99%) and amorphous boron (99%) with the stoichiometry of $MgB_2$ are well mixed. The mixed powder is then loaded into a Fe tube followed by a standard powder-in-tube (PIT) method for the fabrication of the Fe-clad $MgB_2$ wire. The composite tube is drawn from 10 mm to 1-3 mm diameter.

Short length wire samples of about 2 cm in length are sealed in a small Fe tube and then directly heated at a preset temperature ($T_{max}$) for 3-32 minutes in flowing high purity Ar or nitrogen or in air (when a short sintering time is used). This is then followed by a quench to room temperature in air or in liquid nitrogen. Fig. 2 shows the real temperature of the samples as a function of time, starting from when the wires were loaded into a hot tube furnace held at a constant temperature $T_{max}$ of 745, 840, and 900 $^{o}$C. It shows that only a short time (2-3 min.) is required for the samples to reach $T_{max}$, and that the higher the $T_{max}$, the shorter the time. Six samples, which were heat treated at different $T_{max}$, are illustrated and the removal time indicated by open circles as shown in Fig.1. Samples 1, 5 and 6 were removed from the furnace after 3 minutes, having experienced only a few seconds at $T_{max}$. Samples 2-4 were removed after sintering for 6, 15, and 32 minutes. After quenching in liquid nitrogen, the surface of the Fe tube used to seal the wires was slightly oxidised after sintering in air. However, the $MgB_2$/Fe wire samples sealed inside the Fe tube were as fresh as before sintering, regardless of the time at $T_{max}$ and regardless of the atmosphere. A longer sintering time only gives rise to more severe surface oxidation of the outside Fe tube.

## 3. Results and Discussion

XRD patterns were recorded for all the samples. Results show that all the samples have almost the same phase purity (above 90% $MgB_2$). Fig. 2 shows XRD patterns recorded from the core of the Fe-clad MgB2 tapes (samples 1, 4 and 5) after the iron sheath was mechanically removed. It is clear that sample 5 has the same high phase purity as sample 4, even though it was only heat treated for 3 min. as opposed to 32 min.

SEM examination revealed that the grain size is smaller than 1 micron, and the homogeneity appears to be the same for all the samples. An optical image of a cross section of sample 5 is shown in Fig. 3. There is a well-defined interface between the Fe sheath metal and the $MgB_2$ core. It is also noted that the density of the wire sample is only 1.6 g/cm$^3$, suggesting that $J_c$ could be further improved if the density of the wires can be increased.

Due to a strong shielding effect from the Fe sheath metal [9], bare cores were used for the magnetic characterisation. Cylindrical bars of $MgB_2$ core were obtained by removing the Fe sheath mechanically. Transition temperatures $T_c$ and transition widths $\Delta T_c$ for the 3-15 minute treated samples are all very similar. In fact, $T_c$ is almost the same (~ 38 K) for all the samples while there is only a small difference in $\Delta T_c$, which decreases with increasing heating time from 3-15 min for samples 1-3 sintered at 745 $^oC$ (Fig 4).

Measurements of the M-H loops at different temperatures were carried out on these bare cylindrical bar samples. A typical M-H loop of sample 5 is shown in Fig. 5. We can see that a typical flux-jumping pattern is present for temperatures below 15 K. This flux jumping was first observed in an $MgB_2$ bulk samples as reported by our group [14] and also in thin films samples as well [15]. The flux jumping has been directly visualized using magneto-optical imaging techniques and explained in terms of phenomena associated with rapid flux penetration [15].

The critical current density was calculated from the M-H loops using the Bean critical model $J_c = 30\ \Delta M/d$, where $\Delta M$ is the height of the M-H loop, and d is the diameter of the cylinder of the bare core. $J_c$ versus magnetic field up to 6 Tesla for three samples at 10 K, 15 K, 20 K, and 30 K is shown in Fig. 6. It is noted that $J_c$ of $4.5\times10^5$ A/cm$^2$ at 15K and zero field has been achieved for sample 4. Because of the flux jumping the $J_c$ below 15 K cannot be measured. For $T_{max}$ = 745$^o$C, the $J_c$ increases as sintering time increases from 3 to 15 min. A further increase of time up to 32 min. degrades the $J_c$. Fig. 7 shows the $J_c$ versus sintering temperature for samples 1, 5 and 6 which were all treated for 3 min. Sample 1, sintered for 3 minutes at 745 $^o$C, has a markedly lower $J_c$ than the other samples, probably due to poor grain connectivity. However, if $T_{max}$ = 840 $^o$C , the $J_c$ of the wire treated for just 3 minutes (sample 5), is as good as that of a wire treated for 15 min. at 745 $^{o}$C. Furthermore, $J_c$ – field performance of the sample sintered at $T_{max}$ = 840 $^{o}$C for 3 min. is the best out of all the samples, as evidenced by the crossover (indicated by arrows) of $J_c$ – H curves in higher fields as shown in Fig. 6.

Fig. 8 shows the comparison of $J_c$ at 20 K for both zero field and 3 T for sample wires sintered for different times. We can see that the $J_c$ is as high as 3 $\times 10^{5}$ A/cm$^2$ at 20 K zero field for samples 2-5. It is noted that $J_c$ for the 3 min sintered sample 5 is the same as for samples 2-4 which were sintered for 6 min to 32 min. For further comparison, $J_c$ data from a Fe-clad $MgB_2$ tape which was prepared by 3h heating to 800 $^{o}$ C, holding for 1h, and then slow cooling down to room temperature is also shown in Fig.7. It can be seen that the $J_c$ and field dependence of the wire samples which were sintered for only a very short time are as good as for this reference tape. In fact, $J_c$(20K and 3T) for sample 5 is better than for the reference tape and all the other wire samples, indicating that sample 5 has better pinning.

Fig. 9 shows the irreversibility fields ($H_{irr}$) versus temperature for all the samples. $H_{irr}$ was determined from $J_c$ – H curves using the criterion of 100 A/cm$^2$. We can see that all the samples, except sample 1, have approximately the same $H_{irr}$, with sample 5 ( 3 min sintering at 840$^{o}$C) having slightly better $H_{irr}$ .

It should be pointed out that all the $J_c$ data we present here are from wire samples, not from tapes. Further work is needed to fully optimise the condition of this short sintering process. The density of the bare cores for the present wires is only about 1.6 g/cm$^3$. As a result of the fabrication process, tape cores are denser than wires, we can expect that a higher $J_c$ can be achieved for tape samples or higher density wire samples if they also receive the same short heat treatment as that used for these wires.

## 4. Conclusion

In this paper we have investigated the effect of sintering time and temperatures on the formation and critical current densities of Fe-clad $MgB_2$ wires. It was found that there is no need for prolonged heat treatment in the fabrication of Fe-clad $MgB_2$ wires. A total sintering time of several minutes is more than enough to form nearly pure $MgB_2$ with high performance characteristics. The $T_c$, $J_c$ and $H_{irr}$ results show convincingly that the samples which were sintered for 3 minutes are as good as those sintered for longer times. In fact, the $J_c$ field performance for samples sintered for 3 min. at 840 $^oC$, is always at least as good as any other long sintering samples and even slightly better at high fields. $J_c$ of $4.5\times10^5$ A/cm$^2$ in zero field and above $10^5$ A/cm$^2$ in 2T at 15 K have been achieved for the best Fe-clad $MgB_2$ wires. As a result of such a short sintering period there is no need to use high purity argon protection and it is possible to carry out the heat treatment in a much less protective atmosphere or even in air. These findings substantially simplify the fabrication process and reduce the costs for large-scale production of $MgB_2$ wires.

## Acknowledgment

The authors thank Drs T. Silver, M. Ionescu, A. Pan, E.W. Collings and Mr M. Tomsic for their helpful and comments and discussion. This work was supported by funding from the Australian Research Council and the University of Wollongong. S. Soltanian thanks the Department of Physics, University of Kurdistan, Iran for providing financial support for his PhD study at the University of Wollongong.

**Figure Captions:**

Fig.1. The real temperature of the samples as a function of time, after the wires were loaded into a hot tube furnace held at a constant temperature $T_{max.}$

Fig. 2. XRD patterns recorded from the powdered core of the Fe-clad MgB2 wire samples after the iron sheath was mechanically removed.

Fig. 3. SEM image for a typical transverse cross-section for sample 4.

Fig. 4. Temperature dependence of the real part of the ac susceptibility.

Fig.5. M-H loop for sample 5 at different temperatures.

Fig.6. Field dependence of $J_c$ at different temperatures for samples 3, 4 and 5.

Fig.7. $J_c$ vs sintering temperature of $T_{max}$ for samples 1, 5 and 6 which were all sintered for 3 minutes

Fig. 8. Jc as a function of real sintering time at different $T_{max.}$ $J_c$ data (closed circles) for a normally sintered $MgB_2$/Fe tapes is also shown as comparison.

Fig. 9. Irreversibility line for all the samples.

Table.1. Fabrication conditions and $J_c$ for all the samples

| No. | $T_{max}$ | Time* | $J_c$ (A/cm$^2$) | | |
|---|---|---|---|---|---|
| | | | 15 K, 0T | 20K, 0T | 30 K, 0T |
| 1 | 745 $^oC$ | 3 min. | $1.5 \times 10^5$ | $1.1 \times 10^5$ | $3.2 \times 10^4$ |
| 2 | 745 $^oC$ | 6 min | $2.7 \times 10^5$ | $2.7 \times 10^5$ | $6.5 \times 10^4$ |
| 3 | 745 $^oC$ | 15 min | $4.5 \times 10^5$ | $3.5 \times 10^5$ | $1.3 \times 10^5$ |
| 4 | 745 $^oC$ | 32 min | $3.5 \times 10^5$ | $2.8 \times 10^5$ | $9.8 \times 10^4$ |
| 5 | 840 $^oC$ | 3 min | $3.7 \times 10^5$ | $2.9 \times 10^5$ | $1.1 \times 10^5$ |
| 6 | 900 $^oC$ | 3 min | - | $3.0 \times 10^5$ | $1.0 \times 10^5$ |

*Samples were quenched in liquid nitrogen after a total sintering time in a furnace with predetermined $T_{max}$.

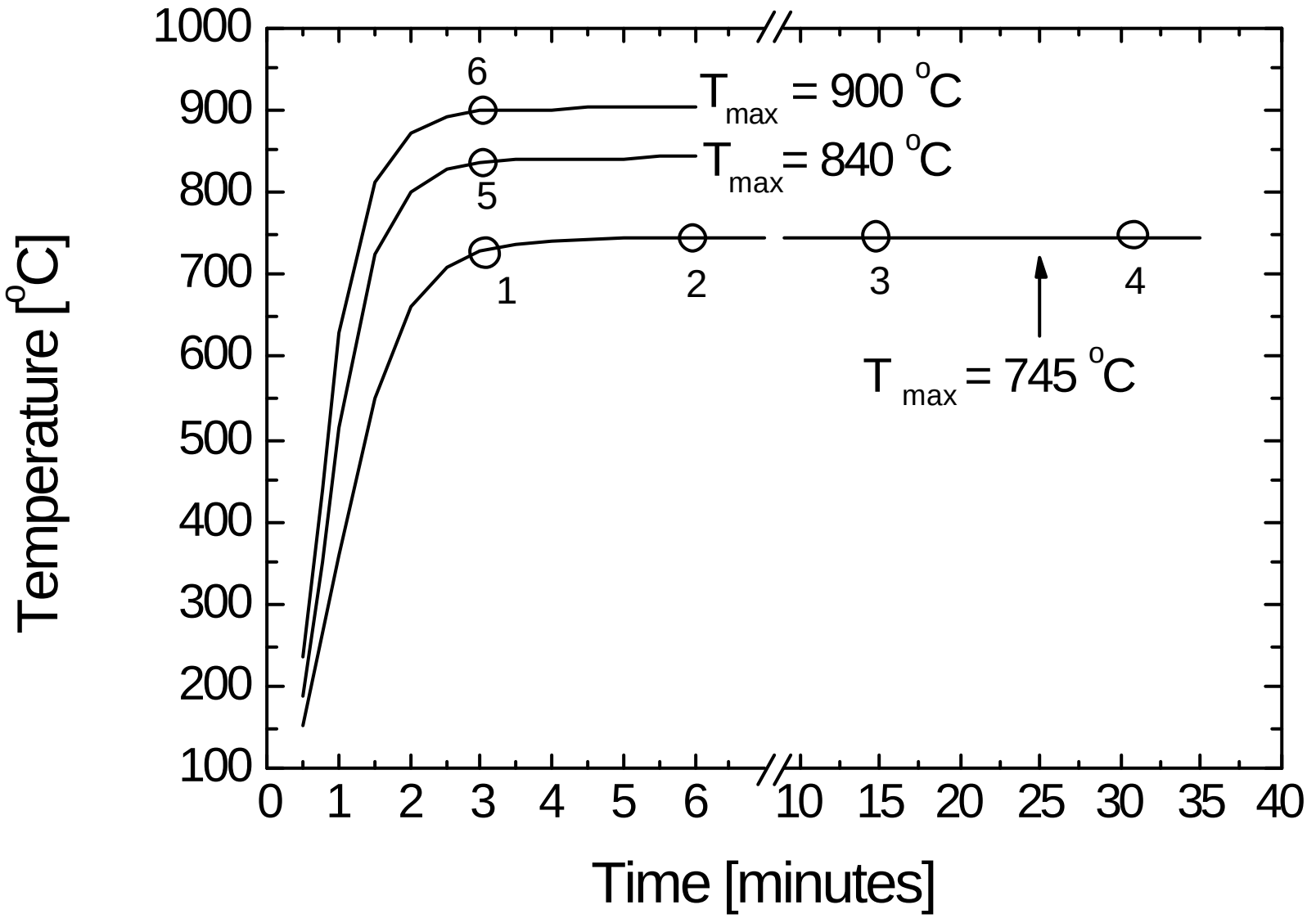

Temperature [°C]
Time [minutes]
1000
900
800
700
600
500
400
300
200
100
0 1 2 3 4 5 6 10 15 20 25 30 35 40
6
5
1
2
3
4
T max = 900 °C
T max = 840 °C
T max = 745 °C

Fig. 1.

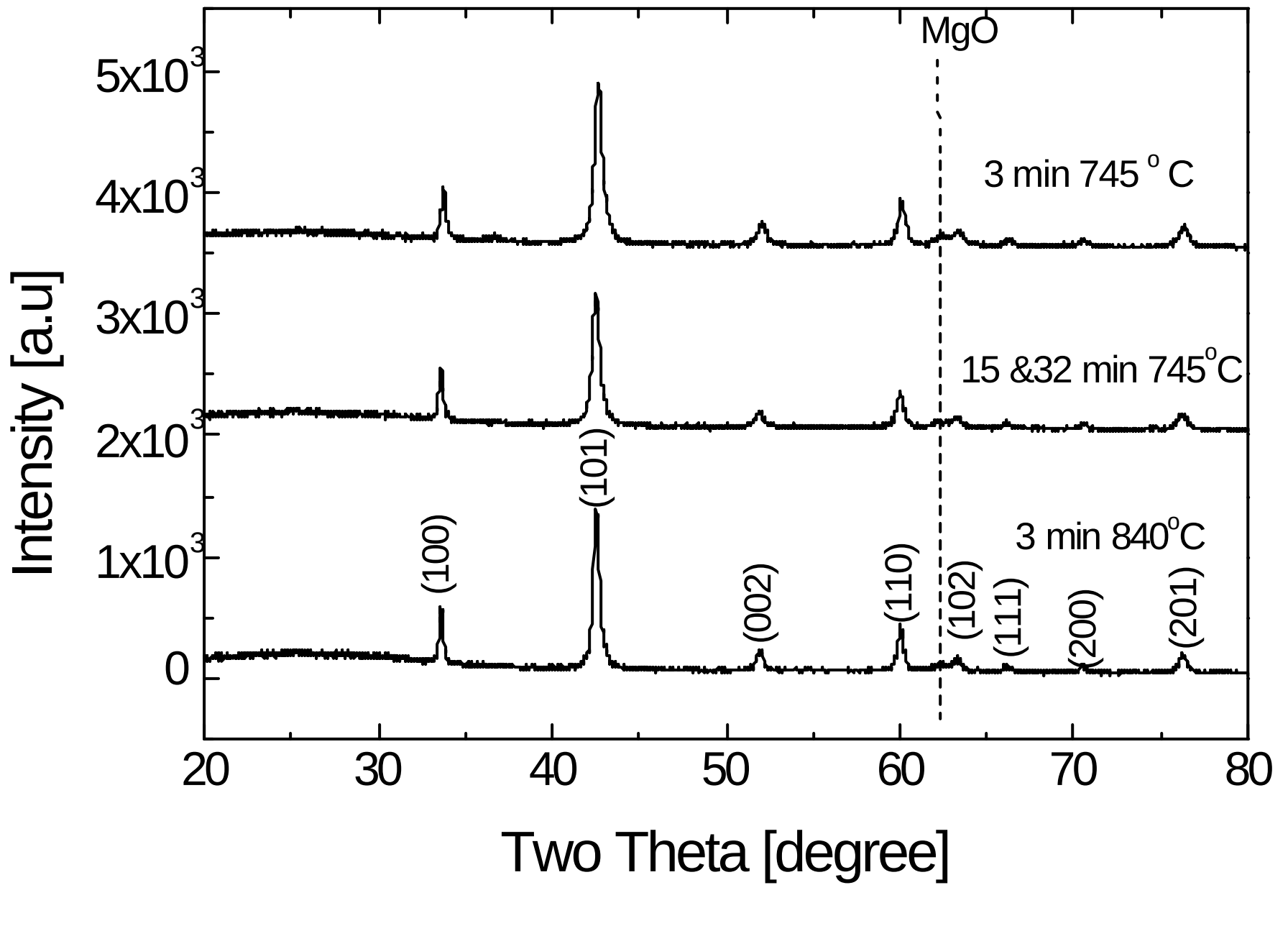

Fig.2.

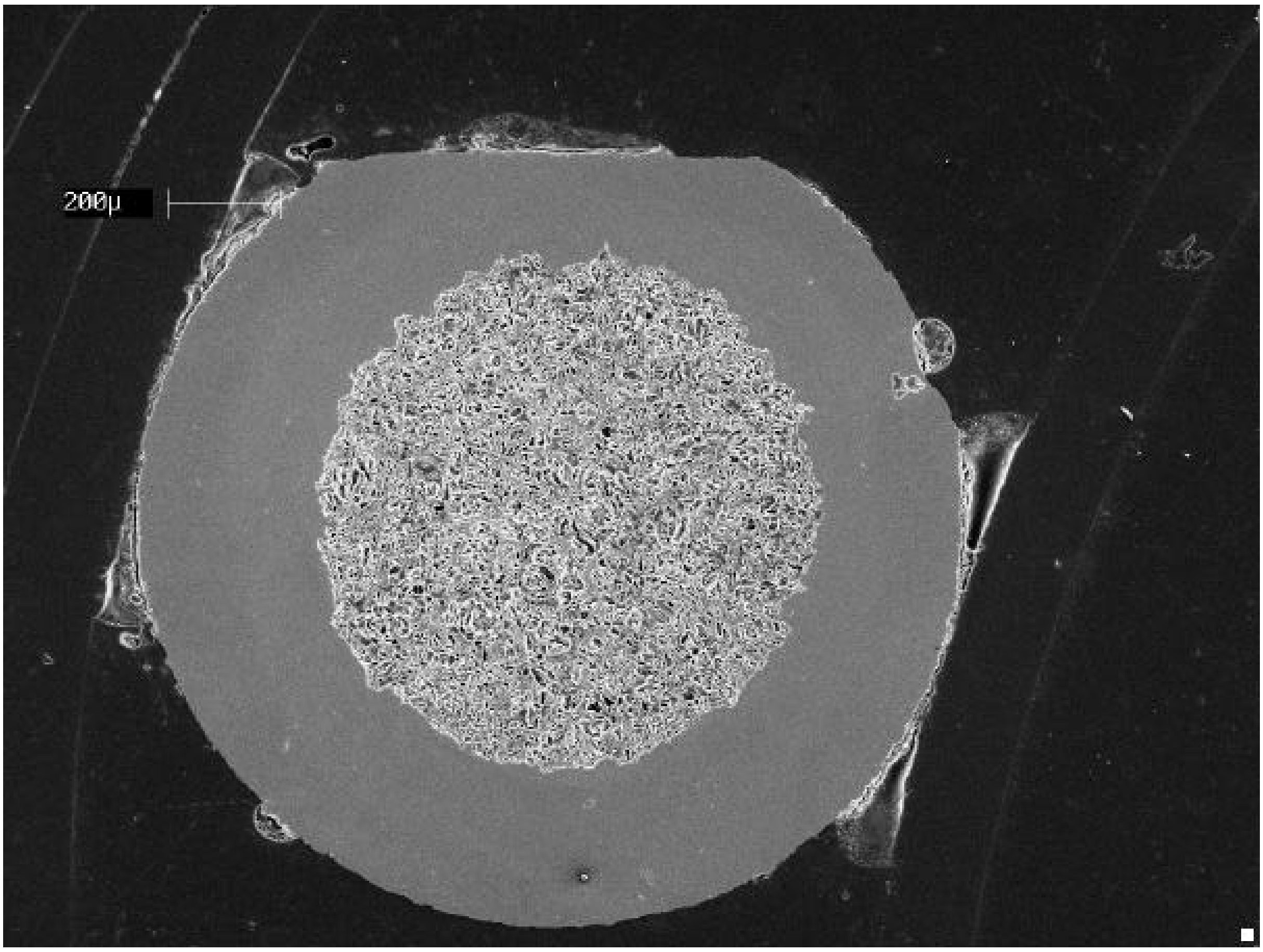

Fig.3.

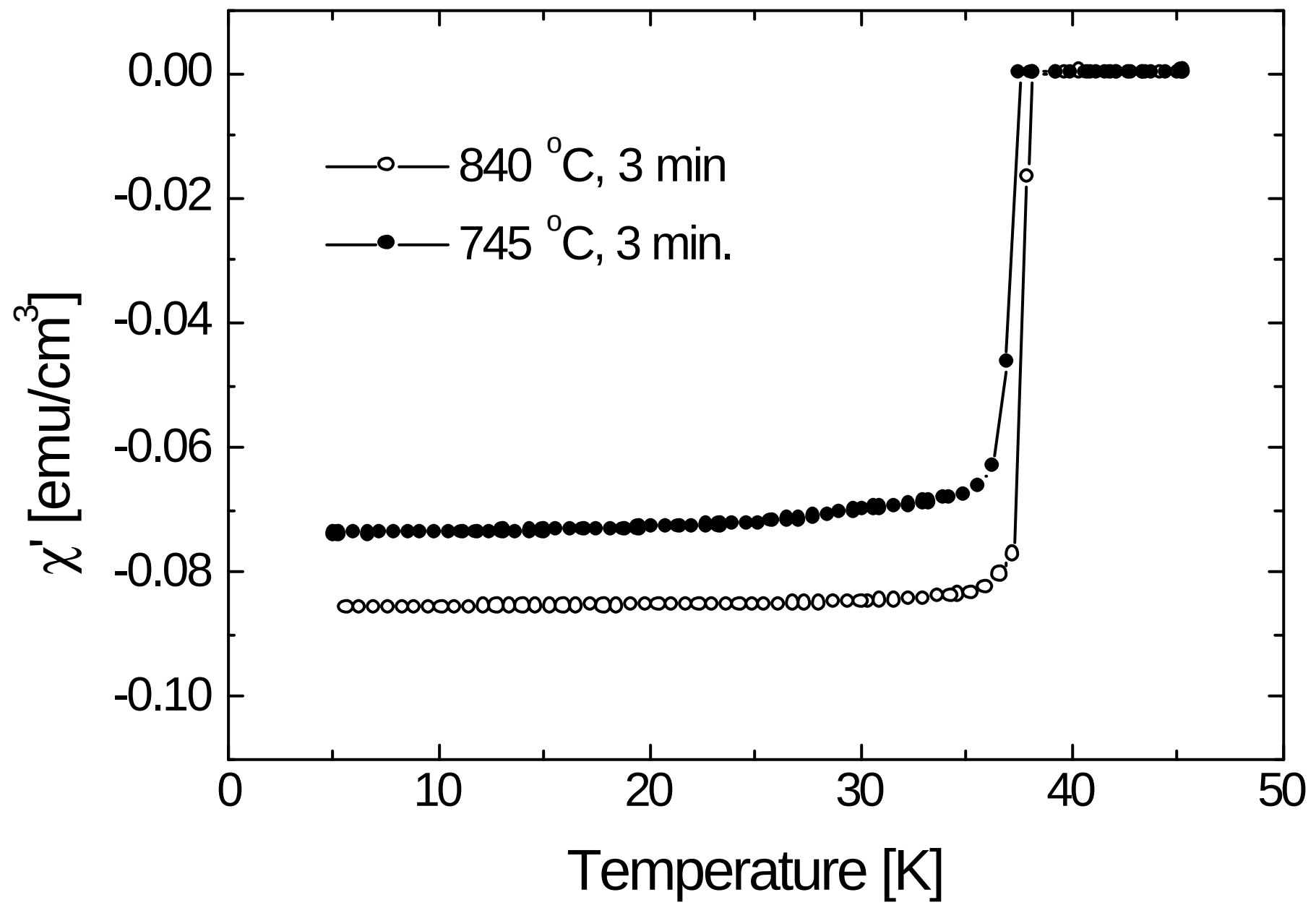

Fig.4.

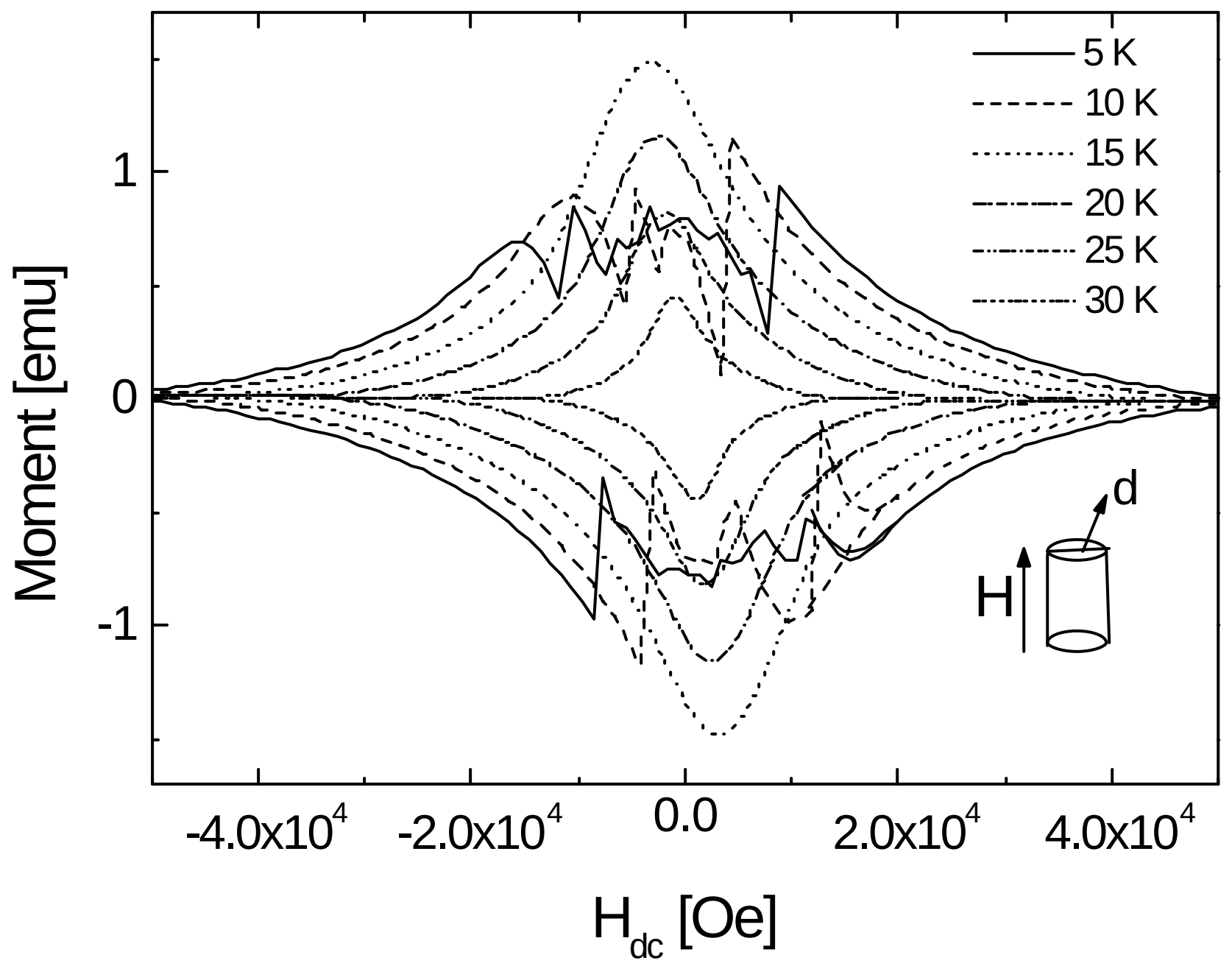

Fig. 5.

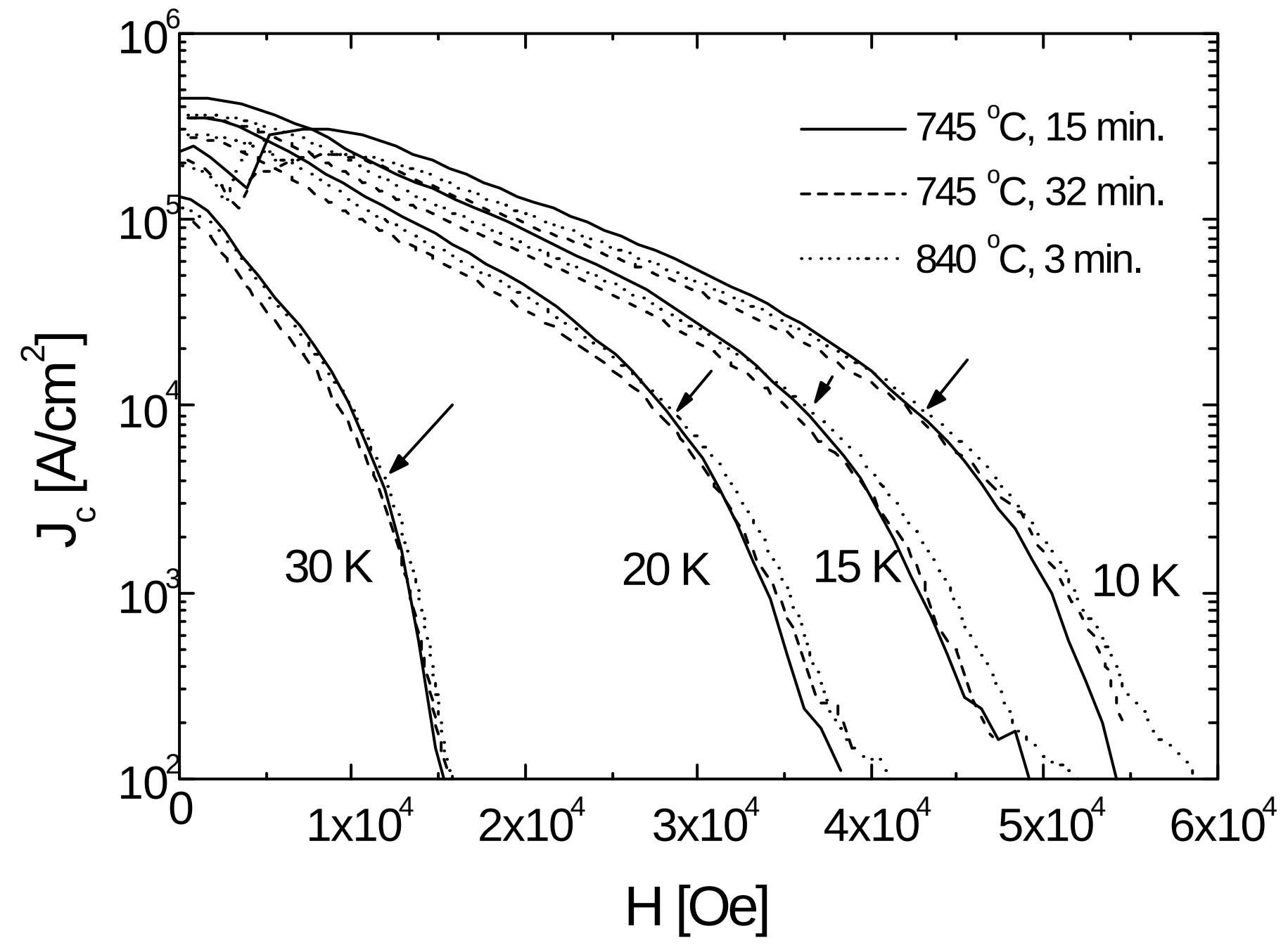

Fig. 6.

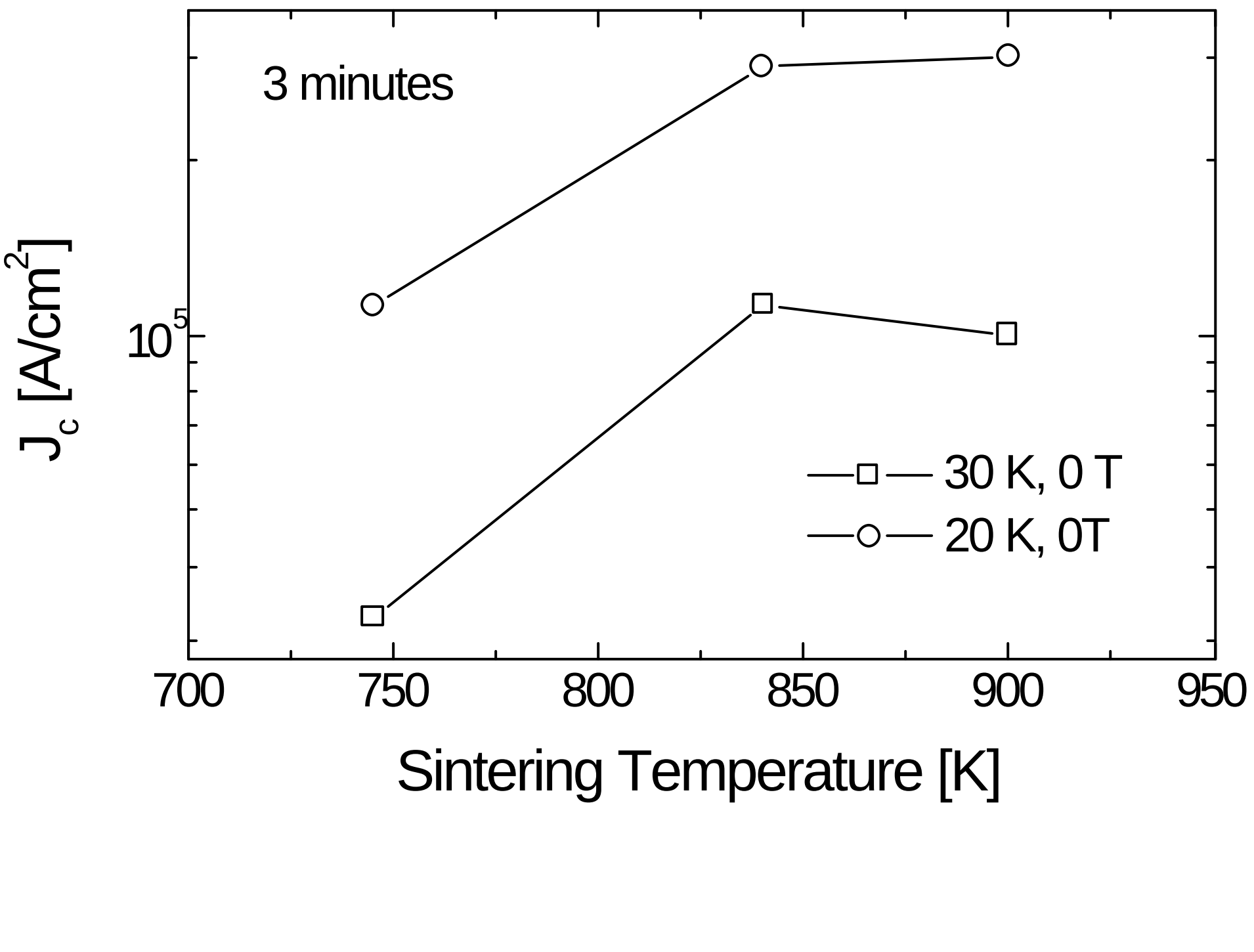

Fig.7.

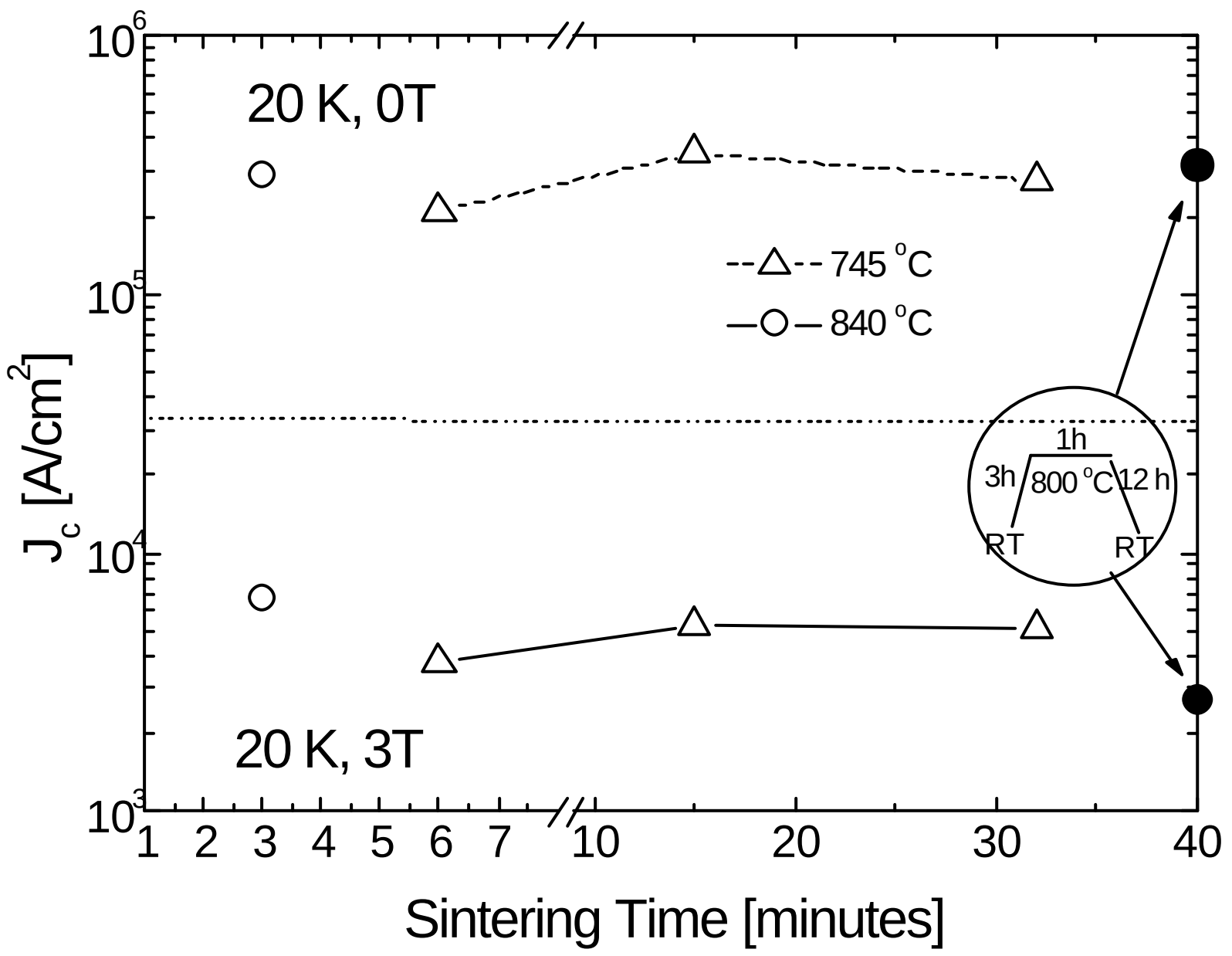

Fig. 8.

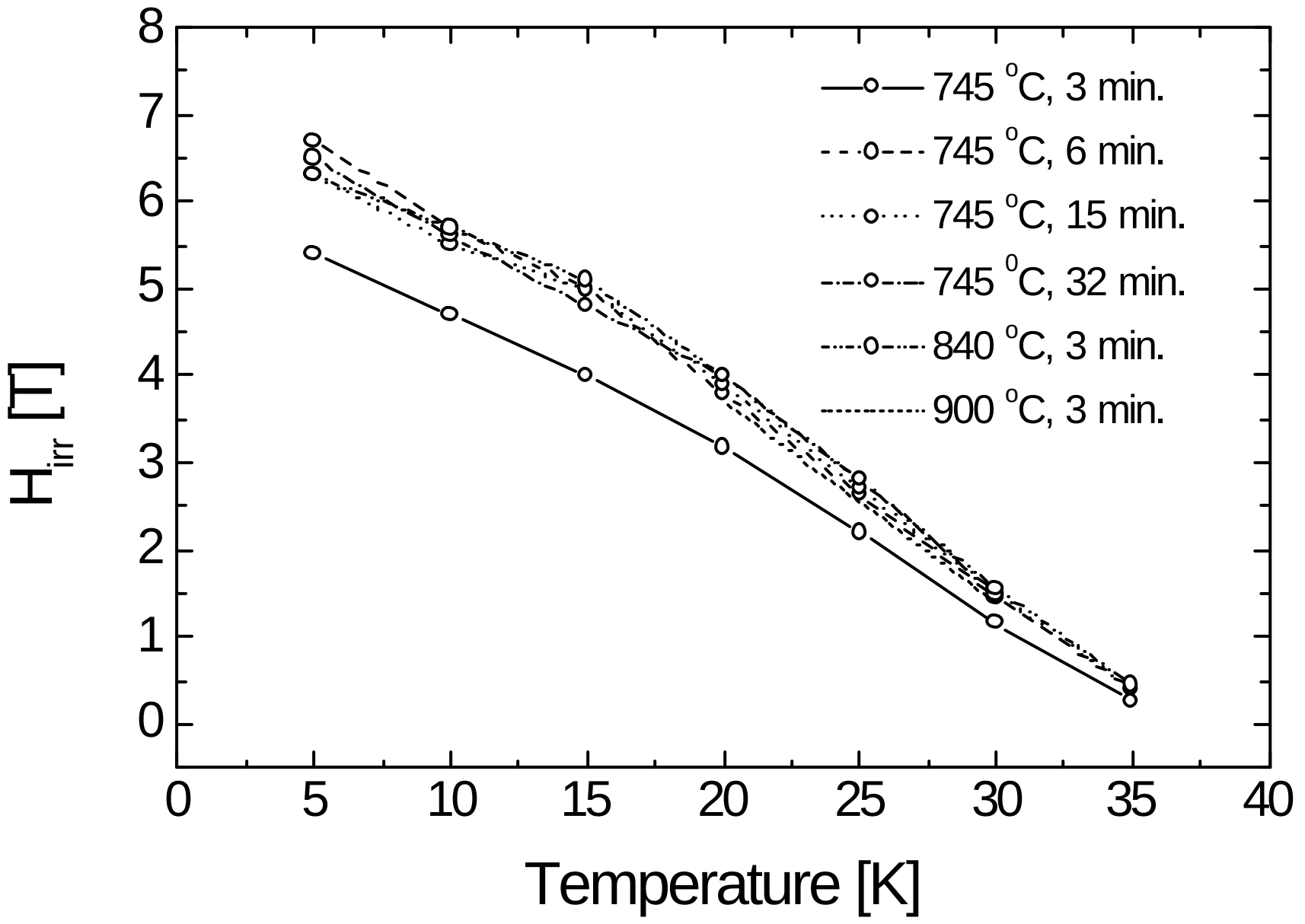
745 °C, 3 min.
745 °C, 6 min.
745 °C, 15 min.
745 °C, 32 min.
840 °C, 3 min.
900 °C, 3 min.
$H_{irr}$ [T]
Temperature [K]

Fig. 9.